\documentclass[11pt, notoc]{JHEP} 
\usepackage{latexsym}
\def\O{\Omega}  
 
\def\bra{\langle}  
\def\ket{\rangle}  
\def\a{\alpha}  
\def\b{\beta} 
\def\g{\gamma}           
  
\def\d{\delta}
\def\D{\Delta}           
\def\e{\epsilon} 
  
\def\et{\eta}  
\def\f{\phi}
 
\def\l{\lambda}         
\def\L{\Lambda}  
  
\def\m{\mu} 
\def\n{\nu}  
\def\s{\sigma}

\def\r{\rho}
\def\th{\theta} 
  
\def\ra{\rightarrow}

\def\Pl{\ell_{\sm{Pl}}} 
\def\dd{\mbox{d}}
\newcommand{\ti}[1]{\tilde{#1}} 
 
\newcommand{\sm}[1]{\mbox{\scriptsize #1}}  
\newcommand{\tn}[1]{\mbox{\tiny #1}} 
\renewcommand{\@}[1]{\sqrt{#1}} 
 
\def\be{\begin{eqnarray}} 
\renewcommand{\le}[1]{\label{#1}\end{eqnarray}}  
\def\ee{\end{eqnarray}} 
\newcommand{\eq}[1]{(\ref{#1})}  
\def\nn{\nonumber\\}

\def\ffract#1#2{\raise .35 em\hbox{$\scriptstyle#1$}\kern-.25em/ 
\kern-.2em\lower .22 em \hbox{$\scriptstyle#2$}}

\def\GN{G_{\mbox{\tn N}}}

\def\pa{\partial}
\def\na{\nabla}
\def\half{{1\over2}}

\title{Boundary description of Planckian scattering in curved spacetimes}
\author{
Giovanni Arcioni\thanks{E-mail: {\tt arcioni@to.infn.it}}\\
Dipartimento di Fisica Teorica, 
Universit$\grave{a}$ di Torino,\\
INFN, Sezione di Torino,\\
Via P. Giuria 1, I-10125 Torino, Italy}
\author{
Sebastian de Haro\thanks{E-mail: {\tt haro@phys.uu.nl}}\\
Spinoza Institute and Institute for Theoretical Physics,\\
Utrecht University\\
Leuvenlaan 4, 3584 CE Utrecht, The Netherlands}
\author{
Martin O'Loughlin\thanks{E-mail: {\tt loughlin@sissa.it}}\\
S.I.S.S.A Scuola Internazionale Superiore di Studi Avanzati\\
Via Beirut 4, 34014 Trieste, Italy}

\abstract{
We show that for an eikonal limit of gravity in a space-time of any 
dimension with a non-vanishing cosmological constant, the Einstein -- 
Hilbert action reduces to a boundary action. This boundary action 
describes the interaction of shock-waves up to the point of evolution 
at which the forward light-cone of a collision meets the boundary of 
the space-time. The conclusions are quite general and in particular 
generalize work of E. and H. Verlinde. The role of the 
off-diagonal Einstein action in removing the bulk part of the action is 
emphasised. We discuss the sense in which our result is a particular
example of holography and also the relation of our solutions
in $AdS$ to those of Horowitz and Itzhaki.}
\preprint{DFTT 7/2001\\SPIN-2001/06\\ITP-UU-01/11\\SISSA 31/2001/EP\\
hep-th/0104039} 
\rm\large 

\begin{document}
\section{Introduction}
Although one could claim that high-energy scattering in gravity should be 
treated in string theory the philosophy adopted in this article, based 
on the holographic principle is that such collisions should be treatable 
in the context of quantum gravity. The holographic principle is taken to 
be the guiding feature behind quantum gravity, rather than the string 
principle. As such it 
implies a reduction in the true number of quantum gravity degrees of 
freedom in line with the counting of degrees of freedom in string theory. 
Thus implementing correctly the holographic principle \cite{ghologr,
g9607,Susskind,AdSreview} in quantum gravity 
should result in a softening of amplitudes akin to that which occurs in 
string theory. From hereon all discussions will take place in the context 
of gravity using the Einstein -- Hilbert action including cosmological 
constant apart from some string-theory related comments in the final
sections. 

The role of high energy scattering has been emphasized by 't Hooft in the 
context of the black hole evaporation process. As is well known, the
appearance of Hawking radiation can be attributed to the enormous red-shift
of outgoing wave packets when propagated back to the region close to
the horizon. Quantum gravitational effects are therefore expected to
play a fundamental role and their inclusion is expected to restore the
unitarity of the Hawking radiation. According to the picture of 
't Hooft 
these gravitational interactions close to the horizon can be effectively
described by shock wave configurations associated to the boosted
particles. They have non-trivial backreaction effects, bringing about
a shift in the geodesics of the outgoing particles and in the position
of the horizon. These correlations should in principle reduce 
the enormous degeneracy of states at the horizon of the black hole that 
one naively calculates using quantum field theory in the curved space-time
of the near-horizon geometry. In this picture
the horizon of the black hole becomes a sort of fluctuacting
membrane due to incoming and outgoing particles and information of the
bulk spacetime is  projected holographically onto this surface. 

In view of these developments it seems interesting to search for a 
more concrete relation between the general arguments of 't Hooft and 
Susskind and the AdS/CFT construction. In this paper we discuss in 
general the eikonal limit of scattering in curved space-times
and find that under certain rather general assumptions about the 
relevant classical backgrounds, the dynamics of gravity are 
described by a theory that lives only on the boundary of the space-time. 
We also find that, from the bulk point of view, some of the classical 
solutions of this boundary theory describe shock-waves moving from the 
boundary to the bulk in Einstein spaces. 
On the way to finding classical backgrounds for our quantum theory and 
checking the classical limit of the theory we need the general solution 
of a two-dimensional gravity model analysed in \cite{BOL}.
Our solutions include the shock-wave solution 
constructed by Horowitz and Itzhaki \cite{HI} and this will be discussed in 
some detail in section 7.

Before proceeding, let us add that even if among our back-ground solutions 
we find black holes, in the eikonal regime one does not strictly
expect black hole formation to appear. To describe the creation of small 
black holes one has to increase the energy transferred during the collisions. 
In other words, one has to go beyond the eikonal approximation. Black hole 
formation is an interesting issue which has been considered in a simplified 
2+1-dimensional set-up in \cite{hans-juergen}. Important related 
discussions in the context of the AdS/CFT correspondence and string 
theory can be found in \cite{BBKVR,Kirill}. 

This paper is organised as follows. In section 2 we will describe the 
setup in which our analysis takes place in particular reviewing the 
basic idea of \cite{VVErice} in which a rescaling is made of the Einstein --
Hilbert action thus separating it into three pieces each scaling differently
in the eikonal limit. In section 3 we discuss the solutions to the classical 
part of this action in various regimes. In section 4 we introduce
shock-wave configurations and then in section 5 we show how the off-diagonal
part of the Einstein equations will be implemented. In section 6 
we discuss the derivation and details of the resulting boundary action
and in section 7 we show how our analysis is related to and extends the 
construction of Horowitz and Itzhaki \cite{HI}. Finally in section 8 
we make some comments on our results and some other concluding remarks. 

\section{The setup}

We consider high-energy scattering in spacetimes with a non vanishing 
cosmological constant $\L$. Our basic construction is a direct 
generalization of that used in \cite{VVErice,VV2} and thus we will consider
an almost forward scattering situation. One introduces two scales, 
${\ell_{\parallel}}$ 
and ${\ell_{\perp}}$: the former
is the typical longitudinal wavelength of the particles while the  latter 
represents the 
impact parameter. Due to the presence of the cosmological constant we also
have an additional
scale ${\ell}$ -- the radius of curvature $\Lambda\sim\frac{1}{\ell^2}$.
For high-energy forward scattering
${\ell_{\parallel}}$ is typically of the order of the Planck length 
$\Pl$, ${\ell_{\perp}} \gg {\ell_{\parallel}}$. This set of 
length scales characterizes the eikonal limit of the scattering
process which for gravity is a linearized regime. We will also deal
with two different cases according to large or small values of the
cosmological constant present in the problem. In general we then 
find that for $\ell_{\perp}$ small on the cosmological scale the 
scattering takes place in the locally flat space-time. On the other
hand for impact parameters that are large on the cosmological scale, 
there are significant changes in the scattering process due to the
curvature. The final result is conceptually the same however as we
find that for shock-wave scattering the process can always be described by 
a lagrangian on the boundary at infinity of the space-time. 

Our general strategy will be to choose dimensionless 
coordinates by extracting the natural length scale in the corresponding 
directions and therefore we will consider the Einstein -- Hilbert 
action plus a non-vanishing cosmological constant and exterior 
curvature $\kappa$,
\be
S={1\over{\Pl}^{d-2}}[\int_M\dd^dx\,\@{-G}\,(R-2\L)] 
+ \frac{1}{\Pl^{(d-2)}}\int_{\partial M}\kappa,
\le{EH}
making a rescaling in the longitudinal $x^\a$  and transverse
coordinates $y^i$ according to the respective scales. 
Under a rescaling of the metric, the action rescales as
\be
\e^{d-4}S_E = \left(\frac{S_0}{\epsilon^2}+\frac{S_1}{\epsilon} 
+ S_2\right)
\ee
where $\epsilon=\Pl/\ell_\perp\sim\ell_\parallel /\ell_\perp$ 
is a very small dimensionless
parameter. $S_2=S_\parallel$ therefore is the strongly coupled part of the
action while $S_0=S_\perp$ is the weakly coupled part. The former is
non pertubative while the latter is essentially classical. The role of 
$S_1$ will be discussed in the following but as is clear it also contributes
to the classical part of the action in the limit of small $\epsilon$.
Under the above rescaling the cosmological term scales as
${\ell^2_\perp\over\e^{d-4}}$ and thus becomes part of the
classical $S_\perp$ or the ``quantum'' $S_\parallel$ depending on the
size of $\ell_\perp$ in comparison to the cosmological scale, $\ell$.
 We will consider
both the case in which the cosmological constant is added to the
classical part of the action -- the ``strongly curved regime'' 
or the regime of 
large impact parameter -- and the case when 
the cosmological constant is included in the strongly coupled part 
of the action -- the ``flat regime'' or regime of small impact parameter. 

\subsection{Scaling and small fluctuations}

We will actually consider a metric that at leading order is block 
diagonal -- the blocks corresponding to the plane of the scattering 
and the plane transverse to the scattering. We will consider a rescaling
of the metric (equivalent but more convenient than that of the 
coordinates discussed above) such that
\be
G_{\mu\nu} = \left(\begin{array}{cc}
g_{\a\b} & h_{\a i}\\    
h_{i \a} & h_{ij}
\end{array}\right) \ra \left(\begin{array}{cc}
\ell_\parallel^2\,g_{\a\b} & \ell_\parallel\ell_\perp\, h_{\a i}\\    
\ell_\parallel\ell_\perp\, h_{i \a} & \ell_\perp^2\, h_{ij}
\end{array}\right).
\le{metric}
Greek indices $\mu,\nu,\dots$ refer to all space-time coordinates whilst 
the indices $\alpha,\beta,\dots$ refer only to the two-coordinates
of the longitudinal scattering plane and the latin indices $i,j,\dots$ 
refer to the directions transverse to the scattering plane. 
In addition to this rescaling of the energy scales, we will also make the 
assumption that the off-diagonal blocks of the metric are 
small. In the end then we will be making a double expansion of the 
action, in $\epsilon$ and in $h_{i\a}$. 

In the limit that $\epsilon\ra 0$ the leading terms in the action 
become classical
and thus we need to derive and examine first the equations of motion
arising from $S_0$ and $S_1$ given our choice of metric. 
$S_0$ always becomes a covariant $1+1$ dimensional 
action and has no terms linear
in the small off-diagonal part of the metric $h_{i\alpha}$.
$S_1$ starts at linear order in $h_{i\alpha}$ and the equation of
motion here comes from the variation with respect to $h_{i\alpha}$
imposing the vanishing of the off-diagonal block of the 
Ricci tensor $R_{i\a}$. The remaining part $S_2$ of the action
is the most interesting part as it is not removed in our
limit and basically describes the dynamics of the eikonal limit
of scattering at high-energy and large impact parameter. 
We will find that this action contains no bulk degrees of 
freedom and thus reduces to a boundary term.
The details of the scaling of the curvature components are in Appendix A. 
Taking into account the fact that $g^{\a\b}$ scales as $\epsilon^{-2}$ 
relative to the scaling of $g^{ij}$, 
the order $\epsilon^{-2}$ term in the action has contributions
from $R_{\a\b}$ at order $\epsilon^0$ and from $R_{ij}$ at
order $\epsilon^{-2}$; the subleading order at $\epsilon^{-1}$ in the action
comes solely from the leading term in $R_{i\a}$; while the final 
term at order $\epsilon^0$ has contributions from the remaining 
lowest order terms in $R_{\a\b}$ and $R_{ij}$.
The resulting double expansion in $\epsilon$ and $h_{i\a}$ is;
\be
\epsilon^{d-4}\,S &=&
{1\over \epsilon^2}\int_{\cal M}\sqrt{-gh}\,\left(R_g + \frac{1}{4}
(h^{ik}h^{lm} - 
h^{il}h^{km})
\partial_\a h_{ik}\partial_\b h_{lm} g^{\a\b}\right)\nn
  &-& {2\over \epsilon} \int_{\cal M} \sqrt{-gh}\,h^{i\alpha} R_{i\alpha}\nn
  &+& \int_{\cal M}\sqrt{-gh}\left(R_h + \frac{1}{4}
(g^{\a\b}g^{\g\d} - g^{\a\g}g^{\b\d})
\partial_i g_{\a\b} \partial_j g_{\g\d}h^{ij}\right)\nn
&+& \int_{\partial{\cal M}} \kappa
 -2 \ell_\perp^2 \int_{\cal M}\sqrt{-gh}\, \Lambda
\le{fullaction}

Considering a path integral for this action we see that the first two terms
become classical as $\epsilon\ra 0$. The cosmological constant can be moved
to different orders of $\epsilon$ depending on it's scaling with
$\ell_\parallel$ or with $\ell_\perp$. 
Physically the mobility of the cosmological constant corresponds to the 
relationship between the scale of curvature of the space-time and the 
impact parameter of the scattering process under consideration. In the 
regime for which the curvature of the space-time does not really
enter into the dicussion we find that the analysis is similar to 
that in flat space, though with corrections to the boundary action coming
from the cosmological constant. In the other regime for which the 
impact parameter is larger than the radius of curvature, the space-time in 
the plane of scattering is curved and the analysis more subtle. The 
result again is that the scattering process can be described by a now
non-quadratic lagrangian that lives on the boundary of the 
space-time. 

The contribution of the exterior curvature will follow the usual 
construction of the Einstein -- Hilbert action. It will split under
rescaling to give contributions to the boundary
in such a way that these boundary terms 
have their usual effect. That is, at the leading ``classical'' orders
they will simply cancel boundary terms that come from integrating by 
parts when varying the action to get the equations of motion. The 
rescaling of the coordinates acts on the exterior curvature part
of the action in such a way that it only contributes to the action 
at order $\epsilon^{-2}$ and $\epsilon^{-1}$ and thus will not
provide any addition to our final boundary action which is at 
order $\epsilon^0$. The details of the 
scaling of the exterior curvature part of the action is given in 
Appendix A.

The general setup that is obtained via this rescaling of the action
by the factor $\epsilon$ (which depends on the energy scales of the problem)
is one in which we have an energy dependent action. This means that 
we are not considering a high energy process in a theory that is already 
defined, but rather we are using the high energy ``eikonal'' limit to
define for us a new action that (hopefully) isolates the degrees of 
freedom that are important for the problem at hand. 
In particular, as we will see from the classical solutions that come
from the small $\epsilon$ limit, the space-time splits 
into a $2+(d-2)$ configuration
in which the two parts are coupled only through the constraint 
that the off-diagonal part of the curvature vanish. The interaction
between the two parts of the space-time - that transversal and that 
longitudinal -- is restricted by $R_{i\a} = 0$. 
Therefore in the case of large cosmological constant although one
may be tempted to interpret this as a limit of small $AdS_d$
it is not. It is more simply a case in which the separation of the 
shock-waves in the transverse part of the space-time is large, 
and the size of the $AdS_2$ in the longitudinal space corresponds to 
a large curvature. However this is not obviously the same
as a scattering in say the context of string theory in an $AdS_d$ with 
large curvature, though it does retain some of the important features.

In the next three sections we will discuss in turn each order in $\epsilon$ of 
this rescaled action.

\section{The solutions}

The geometry in the longitudinal plane of the scattering is determined by the 
saddle point of $S_0$. To find the
general solution we have to  generalize the class of vacuum 
configurations allowing among the
various possibilities the presence of the cosmological constant.
We then need to assume that in general the 
transverse metric depends on the longitudinal 
coordinates  through a warp factor
\be
h_{ij}(x^\mu)=e^{\chi(x,y)}\ti h_{ij}(y).
\le{transvmetric}
On the right hand side of this expression $x$ refers to the coordinates 
$x^\alpha,x^\beta,\dots$ of the scattering
plane while $y$ refers to the transverse coordinates $y^i,y^j,\dots$.
This ansatz can also be used to study radial scattering
situations provided one chooses a time and a radial coordinate in the 
longitudinal directions. 
We will find the classical solutions by substituting this ansatz into the 
action. One could equivalently write down the general equations for the 
classical action and then of course substitute this ansatz to find 
the specific solutions. We will treat the $d=3$ case separately due
to various inconvenient factors of $(d-3)$ in the following general analysis. 

\be
S_0 = 
S_{\perp}=\int_M\@{-g\ti h}\,e^{(d-2)\chi\over 2}\left(R[g]-2\L-{(d-2)(d-3)
\over4}g^{\a\b}\pa_\a\chi\pa_\b\chi\right)
\ee
Making the following field redefinition
\be
\phi(x,y)= (\frac{d-3}{2(d-2)})^{1/2} \exp \left( (\frac{d-2}{4}) \chi(x,y) 
\right)
\ee
one gets
\be
S_{\perp}=-8\int_M\@{-g\ti h} \left(g^{\a\b}\pa_\a\phi\pa_\b
\phi-{(d-2)\over 4(d-3)}\phi^2(R[g]-2\L)\right).
\le{Sperp}
The eikonal limit restricts us to consider the extrema of 
\eq{Sperp}. It is interesting to note that with the assumption 
\eq{transvmetric} the problem is reduced to a general two-dimensional gravity
plus scalar field as studied in
\cite{BOL}. More properly, since the transverse fluctuations are
suppressed in the leading order (in $h_{i\a}$) term of the weakly coupled
action, its explicit expression will not contain, as shown by the
scaling arguments, transverse derivatives. 
Therefore the action still depends on all
four coordinates but the dependence on the transverse directions
is only parametric. 
The equations of motion for the metric and the scalar field $\phi$ are:
\be
\pa_\a\phi\pa_\b\phi-{1\over2}\,g_{\a\b}g^{\g\d}\pa_\g\phi\pa_\d
\phi={(d-2)\over4(d-3)}(\L\,g_{\a\b}\,\phi^2 + 
(g_{\a\b}\Box-\nabla_\a
\nabla_\b)\phi^2)
\le{eomI}
\be
\Box\phi+{(d-2)\over4(d-3)}(R[g]-2\L)\phi=0
\le{eomII}

As proved in the paper \cite{BOL}, all classical solutions have a 
Killing vector that is perpendicular to the curves of constant scalar 
field. Therefore for static configurations with $\Lambda < 0$
we can choose the metric to be of the form,
\be
\dd s^2=-e(x)^2\dd t^2+g(x)^2\dd x^2,
\le{longmetric}
where also
\be
\phi = \phi(x).
\ee
More properly in our case, as we will see below, $e$ and $g$ depend 
on the transverse
coordinates too, since we are considering the two dimensional
longitudinal manifold times the transverse space.

The static configurations are the ones relevant to the case of a negative 
cosmological constant. They have a boundary at spacelike infinity. 
Cosmological solutions relevant to de Sitter space have a spacelike boundary 
and are retrieved from the above solutions either by analytic continuation 
or by introducing an explicit time-dependent ansatz. The metric then becomes
for $\Lambda > 0$, 
\be
\dd s^2=-g(t)^2\dd t^2+e(t)^2\dd x^2,
\ee
where now
\be
\phi = \phi(t).
\ee
These are in fact the type of solutions considered in \cite{BOL}.

\subsection{Large Curvature}

For the case of a negative cosmological constant, the general solution to 
these equations can easily (details in Appendix B) be found and is;
\be
\phi(r)&=&\psi(r)^\gamma\nn
e(r)&=&C\psi(r)^{\gamma\over 4Q}\dot\psi(r)
\le{solution}
where
\be
\psi(r) = Ae^{\@{\lambda\over 4Q\gamma}r} + Be^{-\@{\lambda\over 4Q\gamma}r},
\ee
\be
\dd r = g(x)\dd x,
\ee
and 
\be
\gamma = {4Q\over 1 + 8Q}
\ee
and where  $\l=-{(d-2)\over2(d-3)}\L$, 
$Q={(d-2)\over4(d-3)}$. 

Notice that $A,B$ and $C$ are constant with respect to the 
longitudinal coordinates. However they can have an arbitrary 
dependence on the transverse coordinates $y^i$. Their precise form is 
fixed by imposing opportune boundary conditions depending on the spacetime 
under consideration.

It is also interesting to notice that if one takes either of 
$A$ or $B$ to zero, this two-dimensional metric has constant curvature
and is actually just the metric on $AdS_2$ -- the entire space-time metric
being $AdS_2$ times the $(d-2)$-dimensional  transverse geometry plus 
a warp factor. 

\subsection{Small Curvature}

In the small curvature regime, the cosmological constant term belongs
to the strongly coupled part of the action. The classical action 
that we then have to consider is therefore \eq{Sperp} with $\L=0$. 
However, putting the cosmological constant to zero in the solutions
is a singular limit. It is easy to directly solve for the metric in this 
case and one finds:
\be
\phi(r)&=&(Ar+B)^\gamma\nn
e(r)&=&C(Ar+B)^{\gamma\over 4Q},
\le{case1}

Again, $A,B$ and $C$ are allowed to depend on the transverse 
coordinates. The curvature is
\be
R[g]={16QA^2\over(1+8Q)^2(Ar+B)^2}.
\ee
Notice that it is always positive. In the limit $B\ra\infty$ we recover 
flat space, which was not a solution of the equations of motion in the 
strong regime. In the region $r\ll B$, the space has locally positive 
constant curvature.

Note that there is also a degenerate flat space solution for which 
$\phi$ is constant and
\be
e = Ar + B.
\le{case2}

Even though these solutions could not be obtained directly from those 
with $\Lambda$ non-zero  by  setting $\Lambda$ to zero
they can be obtained as near-horizon limits of those geometries, and 
this exactly corresponds to first shifting the coordinate $r$ and then 
taking the limit of small cosmological constant.

The second case, \eq{case2}, is the near-horizon 
geometry for the solutions of
the previous section, for $B/A>0$. Indeed, a simple coordinate 
transformation brings the near-horizon metric into the 
form of the Rindler metric (see Appendix A). In the 
same way \eq{case1} is the near-horizon geometry in the case $B/A<0$, 
and again a simple coordinate transformation brings it into the form 
of a Rindler type metric with singular horizon (again see Appendix A). 

\subsection{Three-dimensional space-time}

The above formulae are not directly applicable for $d=3$, although 
one can obtain the equations of motion by carefully setting $d=3$
in the above equations. The action in three dimensions is,
\be
S_{\perp}=\int\@{-g\ti h}\,\phi^2(R[g]-2\L),
\ee
where $\phi=e^{\chi/4}$. The equations of motion for the scalar 
field and the metric are \cite{deserjackiw}:
\be
\nabla_\a\nabla_\b\phi^2-g_{\a\b}\nabla^2\phi^2-g_{\a\b}\L\phi^2=0,
\ee
\be
R[g]=2\L,
\ee
and so the space-time always has constant curvature.
It is therefore not surprising that the only solution we find is AdS$_2$. 
These equations are totally symmetric 
under interchanges of $\phi$ and $e$, and under reflections 
$r\rightarrow-r$. Therefore the general solution is
\be
\phi(r)&=&A\,e^{r/2\ell}\nn
e(r)&=&B\,e^{-r/\ell},
\ee
where $\ell$ is the AdS radius. This solution corresponds to pure AdS, 
as expected, with a scalar field $\phi$ that vanishes at the boundary 
and has a singularity at the horizon.

When $\L>0$, we obtain 2-dimensional de Sitter space. 

\section{Gravity at high energy and shock waves}

This section is a necessary digression into the shock-wave solutions
to the classical part of our action. We need to 
understand the 
form of these shock-waves as they will motivate our final 
choice for the metric that we will use in the remaining non-classical
part of our action. The physics in the bulk that can be described 
classically via these shock-waves will then be the physics that is
encoded in the boundary action.

It turns out that scattering at planckian energies is dominated by
the gravitational force. Therefore one should have a complete theory 
of quantum gravity to describe these
processes. However already in the eikonal regime that we are 
considering one can use
semiclassical methods to get useful information.

At leading order gravitational interactions can indeed be described  
by shock wave
configurations -- gravitational waves with a longitudinal impulsive 
profile. Essentially this is the gravitational field surrounding a 
particle whose mass is dominated by kinetic energy therefore representing 
a sort of massless regime of General Relativity 
\cite{AiSe,ACV,SdHJHEP,gnp85,KO1,KO2,Penrose,g87,dmcs,sdasetal}.

Explicit solutions in general spacetimes and their physical 
effects have been described by Dray and 't Hooft \cite{gnp85}, using 
the so called cut and paste method. To fix ideas, suppose we want to 
introduce a shock wave along $u=0$ (in light cone coordinates). To 
this purpose, one first performs a cut
$$
v \rightarrow v+\theta (u)f(x^i)
$$
where $\theta$ is the usual step  function and $f$ is a function only of the  
transverse coordinates $x^i$. 
Then one reabsorbs this shift with an opportune 
coordinate transformation to obtain a metric containing a Dirac delta 
function with support at $u=0$.

To give an example, the gravitational field of a massless particle in flat
spacetime can be described by a metric of the form,
$$
ds^2=-dudv -4p\ln(\mid x^i  \mid^2)\delta(u)  du^2 +dx^idx_i,
$$
where p is the momentum of the massless particle.
The physical effects of such configurations play a crucial role in 
't Hooft's description of the evaporation of a black hole. We refer to
\cite{g9607} for a detailed account and references.

Choosing $x^\mu=(x^+,x^-,x^i)$ and placing the massless particle at
$x^+=x^i=0$, a natural way to rewrite this metric is then 
$$
ds^2=\partial_\alpha X^- dx^\alpha dx^+ +  d^2x^i
$$
with
$$
X^-=x^-+p \,\theta(x^+)\ln (\mid x^i  \mid^2)
$$
This means that if we want to describe the scattering of two high 
energy particles before a collision takes place we must use the 
generalized shockwave configuration with the metric 
in the longitudinal plane being of the form,
$$
ds^2=\partial_\alpha X^a \partial_\beta X^b \eta_{ab}\dd x^\a\dd x^\b,
$$
thus allowing a pair shock-waves of the above type in both $x^+$ and in 
$x^-$.
Here the SO(1,1) $X^a$ vectors can in principle depend on all space-time
coordinates. These are the configurations studied in \cite{VVErice} 
and below we will generalize this construction to include the presence of 
curvature in the longitudinal plane.

\section{The constraint and solution-ansatz}

The second order in our expansion is quite simple. It is 
\be
-\frac{2}{\epsilon}\int\sqrt{-gh}h^{i\alpha}R_{i\alpha}
\le{constraint}
As this is order $\epsilon^{-1}$ we also need to implement the 
corresponding equation of motion (as we did for the leading order
in the previous section). In this order basically the equation of
motion appears as a constraint $R_{i\a} = 0$ on the general solutions. 

Before implementing this constraint we will go back to the construction
of \cite{VVErice} where it is shown how to change variables in 
a way that simplifies the following analysis.
The saddle-point of the transverse part of the
action $S_0$ gives the dominant vacuum
field configurations. In the absence of the cosmological constant
there was only
\be
R[g]&=&0\nn
h_{ij}&=&h_{ij}(y)
\le{cases}
As recounted in the previous section for massless shock-wave configurations 
we will choose a parametrization of the metric via diffeomorphisms
that represents these shock-waves,
\be
g_{\alpha\beta}= \partial_\alpha X^a \partial_\beta X^b 
\eta_{ab}
\ee
where the $X^a(x,y)$ are diffeomorphims which relate $g_{\alpha \beta}$
to the flat metric. Note that they are maps of the two dimensional $x^\alpha$
plane onto itself being however allowed to vary in the transverse
directions and therefore represent transverse coordinate dependent
displacements in the longitudinal coordinates. 
These $X^a$ fields have the appearance of diffeomorphisms in the world-volume 
of the two-dimensional sigma-model and as such would appear to not
introduce any new degrees of freedom. However, in the d-dimensional theory
this is no longer really true as we are not considering the full 
transformation 
of the higher dimensional metric under these transformations. Nevertheless, 
due to the constraint coming from the off-diagonal part of the 
Einstein action we will see that these fields do not contribute additional 
bulk degrees of freedom. 

An intermediate and useful step required to derive the boundary action 
and used to great effect in \cite{VV2} is to express the
strongly coupled action in terms of fields $V_i^\alpha$ defined as
\be
\partial_i X^a =V_i^\alpha \partial_\alpha X^a.
\le{V}
These fields were introduced and motivated physically in terms of fluid 
velocity in \cite{VV2}. In the gravitational setup presented here they 
can be thought of as zweibeins (see also \cite{Kallosh}) 
for a two-dimensional sigma-model describing
the embedding of the scattering plane into the transverse space.
They considerably simplify the action and help to conceptualize our
configuration from the sigma-model point of view.

This definition could also have been motivated by the
simple practical consideration that in order to rewrite the strongly
coupled action as a boundary action one needs to remove derivatives in 
the transverse directions to give one an action that is covariant
in the longitudinal directions. As a consequence one tries to express
every derivative in the transverse directions in terms of a derivative
in the longitudinal ones. This is precisely obtained utilizing this
definition of the $V_i^\alpha$ fields. In this way the indices labelling 
transverse directions act as an internal symmetry of the sigma-model
from the point of view of the longitudinal spacetime. This will 
become clear in the next section where we write the general explicit form 
for the boundary action for all $d \geq 3$ and in both the strong 
and weak curvature regimes.

This construction is basically identical for the more general metrics
considered here. 
As we have seen in Section 4, the conditions
\eq{cases} are too restrictive and one ends up in this general setup 
case with a
family of solutions specified by $g_{\alpha\beta}$ and
$\chi$. A natural generalization of  the above parametrization of
$g_{\alpha\beta}$ is then, 
\be
g_{\alpha\beta}=e^{\sigma(X)} \partial_\alpha X^a \partial_\beta X^b 
\eta_{ab}.
\le{anstzI}
thus allowing the presence of a warp factor in the $2+(d-2)$ decomposition
of the metric. 
In principle $\sigma$ may also have some explicit $y$ dependence, however this
would correspond to a more complicated sigma model than the one we
are presently considering. As stated several times, 
the introduction of $X$ is simply
a statement that the scattering configuration described by shock-waves 
is described simply via singular co-ordinate tranformations with 
support only along light-cones in the scattering plane and thus the classical 
solution $\sigma(x)$ after the shock wave ansatz becomes simply
$\sigma(X)$ and similarly $\chi(x)$ becomes $\chi(X)$. 
As in the previous case \cite{VV2}, we define fields $V_i^\a$ by
\be
\pa_iX^a=V_i^\a\pa_\a X^a,
\le{Vfield}
which in turn gives when lowering the longitudinal index
\be
V_{i\a}=e^{\s(X)}\pa_iX^a\pa_\a X_a.
\ee
With the use of the $V_{i\a}$ the longitudinal metric 
changes under reparametrisations of the transverse coordinates 
according to,
\be
\pa_ig_{\a\b}=\nabla_\a V_{\b i}+\nabla_\b V_{\a i}.
\le{cmode}
Finally we also will have 
\be
h_{ij} = e^{\chi(X)} \tilde{h}_{ij}(y).
\le{anstzII}

Notice that this form of the solutions captures both the cases $\L<0$ 
and $\L>0$.

\section{The Effective Boundary Theory}

We now examine how our classical solution -- ansatz leads us to the general
result that in this setup the transverse action $S_\parallel$ always
reduces to a boundary action. 

We now perform the substitution of our solution -- ansatz 
into the leading order ($\epsilon^0$) action,
\be
S_\parallel[g,h]&=&\int\@{-gh}\left[R[h] -{1\over4}\,h^{ij}\pa_ig_{\a\b}
\pa_jg_{\g\d}\,\e^{\a\g}\e^{\b\d}\right]\nn
&=&\int\@{-gh}\left[R[h] -\e_{\a\g}\e_{\b\d}\,h^{ij}\na^\a V^\b_i\na^\g 
V_j^\d +\half h^{ij}R_iR_j \right],
\ee
where
\be
R_i&=&\e^{\a\b}\na_\a V_{i\b}.
\ee

\subsection{Strong curvature regime}

Filling in the solutions of the classical equations of motion,
\be
g_{\a\b}&=&e^{\s(X)}\,\pa_\a X^a\pa_\b X^b\,\et_{ab}\nn
h_{ij}&=&e^{\chi(X)}\,\ti h_{ij},
\ee
we get
\be
S_\parallel&=&\int\@{-g\ti h}\,e^{{d-4\over2}\,\chi}\left[R[\ti h] -
\e_{\a\g}\e_{\b\d} \ti h^{ij}\na^\a V^\b_i\na^\g V_j^\d +\half R_i^2+
\right.\nn
&-&\left.(d-3)\Box_{\ti h}\chi -{1\over4}(d-3)(d-4)(\pa_i\chi)^2\right],
\ee
and from now on we raise and lower transverse indices by means of 
the rescaled metric $\ti h_{ij}$.

The action splits into a bulk and a boundary term:
\be
S_\parallel&=&S_{\sm{bulk}}+S_{\sm{bdry}}\nn
&=&\int_{\pa M}\dd x^\a\@{\ti h}\,e^{{d-4\over2}\,\chi} \left(R[\ti h]\,
e^\sigma \,X^0\pa_\a X^1 +e^\s\e_{ab}\,\frac{}{}
\pa_iX^a\times\right.\nn
&\times&\left.[\pa^i\pa_\a X^b +\half\na^\b\s(\pa_\a X^bV^i_\b -\pa_\b 
X^bV^i_\a)] -\half\,V_{i\a}R^i -{d-3\over2}\,\e_{\a\b}
V^{i\b}\pa_i\chi\right) +\nn
&+&\int_M\@{-g\ti h}\,e^{{d-4\over2}\,\chi}\,V^{i\a}R_{i\a},
\le{boundaryaction}
where by $X^a$ we mean the variation of $X^a$ around its infinite value. 
Filling in the constraint
\be
R_{i\a}&=&\half\,R[g]V_{i\a} +\half\,\e_{\a\b}\na^\b R_i +
\half\pa_\a\chi\na^\b V_{i\b} -{d-3\over2}\,\pa_i\pa_\a\chi +\nn
&+&{d-4\over4}\,R_i\,\na^\b\chi(\na_\a V_{i\b} + \na_\b V_{i\a}) =0,
\le{constrainteq}
it obviously reduces to a boundary action. Note that this action will 
generally consist of two disconnected pieces corresponding to the 
two boundaries of the longitudinal space-time. 

When $\L<0$, as we have been implicitly assuming in this section, the 
boundary is timelike. In the large curvature regime, the discussion for 
$\L>0$ is more intricate. Formally, the above derived action is valid in 
de Sitter space as well, the boundary being now a spacelike boundary at 
the future and past infinities. In this case the 
boundary theory is defined on an Euclidean manifold $\pa M$ and thus
the physical interpretation in terms of causality 
and locality of a corresponding  holographic map is somewhat 
more mysterious than in the AdS case. 
Our derivation suggests that a certain set of observables 
in a holographic description of 
de Sitter space can be defined as correlation functions of a theory
living on a space-like surface\cite{Bousso}

\subsection{Weak curvature regime}

In this regime there are  two types of solutions, curved (singular) and 
flat. The action is the same as in the strong curvature regime, apart
from an additional term proportional to the cosmological constant,
\be
S_\parallel[g,h]&=&\int\@{-gh}\left[R[h] -2\L -{1\over4}\,h^{ij}
\pa_ig_{\a\b}\pa_jg_{\g\d}\,\e^{\a\g}\e^{\b\d}\right]\nn
&=&\int\@{-gh}\left[R[h] -2\L-\e_{\a\g}\e_{\b\d}\,h^{ij}\na^\a 
V^\b_i\na^\g V_j^\d +\half h^{ij}R_iR_j \right].
\ee
Again filling in the general solutions we get:
\be
S_\parallel&=&S_{\sm{bulk}}+S_{\sm{bdry}}\nn
&=&\int_{\pa M}\dd x^\a\@{\ti h}\,\left((R[\ti h]-2 e^\chi\L)\,\frac{}{} 
\,X^0\pa_\a X^1 +e^{{d-4\over2}\,\chi+\s}\e_{ab}\,\pa_iX^a\times\right.\nn
&\times&\left.[\pa^i\pa_\a X^b +\half\na^\b\s(\pa_\a X^bV^i_\b -
\pa_\b X^bV^i_\a)] -e^{{d-4\over2}\,\chi}(\half\,V_{i\a}R^i +{d-3\over2}
\,\e_{\a\b}V^{i\b}\pa_i\chi)\right) +\nn
&+&\int_M\@{-g\ti h}\,e^{{d-4\over2}\,\chi}\,V^{i\a}R_{i\a}
\ee

The most interesting case is the flat-space solution, where the action 
is simply quadratic:
\be
S_\parallel&=& S_{\sm{bulk}}+S_{\sm{bdry}}\nn
&=&\int_{\pa M}\dd x^\a\@{\ti h}\,[\e_{ab}\,X^a(\half R[\ti h]-\L -
\triangle_{\ti h})\, \pa_\a X^b -\half\,V_{i\a}R^i] \nn
&+&\int_M\@{-g\ti h}\,e^{{d-4\over2}\,\chi}\,V^{i\a}R_{i\a}
\ee
The constraint then reads
\be
R_{i\a}=\half\,\e_{\a\b}\na^\b R_i=0.
\ee
Note that unlike \cite{VVErice} this does not imply that $R_i = 0$ but
that $R_i$ is a function only of the transverse coordinates $R_i(y)$. In 
particular then even for the flat space we have found that the complete 
analysis of this limit actually implies that there can be an additional 
term in the boundary action. It would be interesting to understand the
physical meaning of this extra piece. 

The action can be rewritten as
\be
S_\parallel&=&S_{\sm{bdry}}\nn
&=&\int_{\pa M}\dd x^\a\@{\ti h}\,\{\e_{ab}\,\pa_\a X^a (\triangle_{\ti h} 
+\L -\half R[\ti h])\,X^b + \half\partial_\a X^aR^i(y) \partial_i
X_a\}
\le{smallcurv}
which will be convenient for the discussions in the next section. 
Needless to say that in this case the classical solutions are independent 
of the value of the cosmological constant, and therefore the action 
\eq{smallcurv} allows any value of $\L$. 
We thus find ourselves with a quadratic action like that of \cite{VVErice}.
Correspondingly there will be a way to quantize this action, write down
the S-matrix and to study the inevitable non-commutativity of the boundary 
coordinates.  In the next section
we will consider the relationship between our construction and the 
curved space-time shock-wave scattering considered in particular in 
the paper of Horowitz and Itzhaki \cite{HI}. 

\section{Shock-waves from eikonal gravity and the AdS/CFT Correspondence}

The boundary action found in the small curvature regime for $R_i(y) = 0 $  
is quadratic and therefore easy to deal with. 
In fact it is a straightforward generalisation 
of the boundary action found in \cite{VVErice,VV2}. 

Let us briefly discuss its quantum mechanical properties when we couple it 
to point particles. In this regime, and restricting ourselves to the 
classical solutions of the equations of motion, the longitudinal space is 
flat. Therefore, the coupling to point particles in this case 
goes precisely along the lines of section 5.1 of \cite{VV2}. Hence we will 
not discuss this issue at length here. For details about the stress-energy 
tensor of a pointlike particle we refer the reader 
to Appendix \ref{nullgeodesics}.

Quantisation of the boundary action in the weak-coupling regime is 
straightforward and, as discussed in \cite{VV2}, it leads to non-trivial 
commutators for the coordinates $X^a$:
\be
[X^a(y),X^b(y')]=i\e^{ab}f(y,y'),
\le{XaXb}
where $f$ satisfies the Green's function equation
\be
(\triangle_{\ti h}+\L -\half\,R[\ti h])\,f(y,y')=\delta^{(d-2)}(y-y').
\le{greeneq}

As we have already discussed we expect shock-waves to also be described by our 
boundary action. Let us first briefly discuss how shock-waves can be 
implemented in AdS \cite{HI}.

We write pure AdS in the following co-ordinates,
\be
\dd s^2={4\over(1-y^2/\ell^2)^2}\,\et_{\m\n}\dd y^\m\dd y^\n,
\le{pureAdS}
where $y^2=\et_{\m\n}y^\m y^\n$. The stress tensor of a massless particle 
is computed in Appendix C and gives 
\be
T_{uu}=-p\,\d(u)\d(\r),
\ee
where $\r$ is the radial co-ordinate $\r=\sum_{i=1}^{d-2}y_i^2$.

Horowitz and Itzhaki found the following solution of Einstein's equations 
with a massless particle:
\be
\dd s^2={4\over(1-y^2/\ell^2)^2}\,\left(\et_{\m\n}\dd y^\m\dd y^\n +
8\pi\GN\,p \d(u)(1-\r^2/\ell^2)f(\r)\dd u^2\right)
\le{HI}
provided
\be
\triangle_h f -4\,{d-2\over\ell^2}f=\d(\r).
\le{shiftads}
$\triangle_h$ is the Laplacian on the transverse hyperbolic space,
\be
\dd s^2={\dd\r^2+\r^2\dd\O^2_{d-3}\over(1-\r^2/\ell^2)^2},
\ee
and therefore \eq{shiftads} takes the form
\be
f''+{d-3+(d-5)\rho^2/\ell^2\over\rho(1-\rho^2/\ell^2)}f'
-{4(d-2)\over\ell^2(1-\rho^2/\ell^2)^2}\,f&=&\d(\rho).
\ee

The solutions to \eq{shiftads} are given by:
\be
{\mbox{AdS}_3:}\,\,\,\,\,\,\,\,\,\,\, f(\r)&=&{\ell\over2}
(C+\th(\r))\,\sinh\log\left({\ell+\r\over\ell-\r}\right) +
\ell D\cosh\log\left({\ell+\r\over\ell-\r}\right)\nn
{\mbox{AdS}_4:}\,\,\,\,\,\,\,\,\,\,\, f(\r)&=&C\,{1+\r^2/\ell^2\over1-
\r^2/\ell^2}\,\log(\r/D)+{2C\over1-\r^2/\ell^2}\nn
{\mbox{AdS}_5:}\,\,\,\,\,\,\,\,\,\,\, f(\r)&=&{C\over1-\r^2/\ell^2}
\left({1\over\r}+{6\r\over\ell^2} +{\r^3\over\ell^4}\right) +
{D\over\ell}\,{1+\r^2/\ell^2\over1-\r^2/\ell^2},
\le{fsol}
$D$ is an arbitrary constant to be determined by boundary conditions. 
$C$ is a constant of order 1 that can be computed either by explicit 
computation or by matching with the Minkowski solutions.
The shift function $f$ of course behaves like the solutions 
for shock-wave in Minkowski space $f\sim{1\over|x|^{d-3}}$ 
in the limit when the AdS radius divided by 
the impact parameter goes to infinity, $\ell/\r\rightarrow\infty$.
In fact the metric \eq{HI} was derived by boosting a black hole to the 
speed of 
light while sending its mass to zero and keeping its energy fixed. 

Notice that for an Einstein space with negative curvature and curvature 
radius $\L=-{(d-1)(d-2)\over2\ell^2}$ the above general equation for the 
shift function derived via our boundary action method \eq{greeneq} 
reduces to the condition \eq{shiftads} 
found by Horowitz and Itzhaki precisely when the transverse space is
Euclidean AdS$_{d-2}$:
\be
\dd s^2=4\,{\dd\r^2+\r^2\dd\O_{d-3}^2\over(1-{\r^2\over\ell^2})^2}.
\ee
In this case, the transverse curvature is
\be
R[\ti h]=-{(d-2)(d-3)\over\ell^2}.
\ee

The class of solutions to \eq{greeneq} is 
however much larger than only shock-waves in pure AdS. It allows for 
solutions where the transverse curvature is postitive, negative or zero, 
and the cosmological constant is also allowed to take 
positive values. In the limit $\L\rightarrow0$, all our results of course 
agree with the results found in \cite{gnp85}.

It is not surprising that we find an approximate shock-wave from our 
boundary action only in the small curvature regime. 
These shock-waves have a smooth 
limit as $\Lambda\rightarrow0$ which of course could not happen in 
the large curvature regime. 

Horowitz and Itzhaki have argued that the CFT duals of shock-waves are 
``light-cone states'' -- states with their energy-momentum tensor 
localised on the boundary light-cone. It is tempting to argue that our 
boundary description should somehow be related to these light-cone states. 
Indeed, we have shown that our boundary theory describes bulk shock-waves 
in an approximate fashion. Hence one is 
led to speculate that our boundary action is somehow related to some sort 
of eikonal limit of a boundary CFT perturbed by the addition of light-cone 
states. Notice, however, that it is not at all clear how to prove such a 
relation.  In particular it is not clear how light-cone states 
should be precisely described in field theory, although some attempts have 
been made in \cite{PST}. Related discussions can be found in 
\cite{PolS-m,SussS-m,Steve} and, recently, in \cite{Steve2}. Furthermore
if quantum gravity has a boundary description at all energies, we have 
nevertheless taken the eikonal limit of it
thereby explicitly breaking the possible covariance of  
the boundary theory. An interesting question is whether it is possible to 
do an eikonal approximation in a covariant way, or whether it is possible 
to restore covariance afterwards. Progress in the latter direction for the 
Minkowski case is found in \cite{g9607,g9805,SdHJHEP}. In particular, in
the simplified 2+1 -- dimensional setup, restoring Lorentz covariance is 
tantamount to going beyond the extreme eikonal regime \cite{SdHJHEP}.

It would be extremely interesting if we could find an analog of \eq{XaXb} 
in the context of the AdS/CFT correspondence. 
This would amount to identifying 
the operators $X^a$ in the CFT and to interpreting them from the bulk point of 
view. Based on previous considerations by 't Hooft and a computation of 
the trajectories of massless particles outlined in Appendix 
\ref{nullgeodesics}, they 
are expected to correspond to the positions of colliding particles, however
a careful analysis is required in order to prove this. This could most easily
be done using the techniques in \cite{KSS1} where boundary sources and 
operators are related to the coefficients of the perturbative expansion of 
bulk fields.

\subsection{Boundary description of scalar fields in an AdS-shock-wave 
background}

So far we have discussed single particles in an AdS background and 
interactions between quantum mechanical particles by means of shock-waves. 
One would however be ultimately interested 
in considering second quantised fields that interact gravitationally. 
In flat space, 
computing an S-matrix and extracting from it the amplitude for scattering 
between massless particles is a relatively straightforward task even if 
the interactions are gravitational \cite{g87,g9607}. In AdS, however, 
things are much more complicated due to the presence of the timelike 
boundary and the impossibility to separate wavepackets. These problems 
can be sidestepped by imposing appropriate boundary conditions on the 
fields and ensuring that the S-matrix is unitary \cite{AvIsSt}. However, 
this is not possible for all the modes, and in the context of the AdS/CFT 
correspondence we are interested in considering both normalisable and 
non-normalisable modes. For other discussions of the AdS S-matrix, see 
\cite{PolS-m,SussS-m,Steve}. In this paper we will not consider this issue, 
but rather concentrate on the CFT duals of scalar fields with generic 
boundary conditions.

As a first step towards considering the full quantum mechanics of scalar 
fields interacting gravitationally in AdS, we consider scalar fields on an 
AdS-shock-wave background. We concentrate on conformally coupled scalar 
fields. These have the nice property that their equation of motion is 
invariant under Weyl rescalings, up to a certain weight. The Klein-Gordon 
equation for these fields is
\be
\left(\Box_G -{d-2\over4(d-1)}\,R[G]\right)\f(y)=0,
\le{37}
and so in the AdS-shock-wave background they have mass $m^2=-
{d^2-1\over4\ell^2}$. Note that in this subsection we use the convention
of the $AdS/CFT$ correspondence for which the boundary of the space-time 
is d-dimensional.

We perform a conformal transformation by which we remove the double pole 
of the metric:
\be
\dd s^2=G_{\m\n}\,\dd y^\m\dd y^\n ={1\over\O(y)^2}\,\bar 
G_{\m\n}\dd y^\m\dd y^\n,
\ee
$G$ being the AdS-shocwkave metric \eq{HI}. The Klein-Gordon equation 
transforms into
\be
\stackrel{-}{\Box}\! \bar\phi(y)=0,
\le{38}
calculated in the metric $\bar G_{\m\n}$, and 
\be
\bar\f(y)=\O^{1-d\over2}\,\,\f(y).
\ee
There is no curvature term in \eq{38} because $R[\bar G]=0$. For the 
metric $\bar g_{\m\n}$, the Laplacian factorises into a flat piece plus 
a shock-wave part,
\be
\stackrel{-}{\Box}\!\bar\f(y)=\et^{\m\n}\pa_\m\pa_\n\bar\f(y) -p\,\d(u)
\,F(\rho)\,\pa_v^2\bar\f(y)=0.
\le{40}
Equation \eq{40} is difficult to solve in general due to the transverse 
derivatives but it 
can be readily solved in the eikonal approximation. A simple plane-wave 
solution is given by 
\be
\phi(y)&=&\O(y)^{{d-1\over2}}\,\exp \left[ikv+ ikp\,\th(u)\,F(\rho)\right],
\le{44}
as one would expect from a computation of trajectories: the only effect 
of the shock-wave is a shift of the wave function over a distance given 
by the shift. The full solution gives, in the eikonal approximation,
\be
\phi(y)=\O(y)^{d-1\over2}\int\dd^dk\, a(k)\, e^{ipk\,\th(u)F(\rho) +
ik_\m y^\m} +\mbox{c.c.},
\le{45}
where $k_\m^2=0$. Note that this sense of eikonal approximation is 
the same as in previous sections -- all transverse derivatives
are set to zero. 

To interpret this classical field from the CFT point of view, it is 
easiest to go to Poincare co-ordinates where the boundary is at $r=0$. 
The above field then has the following expansion \cite{KSS1}
\be\label{expansion}
\phi(r,x)=r^{d-1\over2}\phi_0(x)+\cdots
\ee
as it approaches the boundary. This is the expected behaviour for a field 
of mass $m^2=-{d^2-1\over4\ell^2}$ 
interpreted as the source or as the dual operator.

For $-d^2/4<m^2<-d^2/4+1$, which includes the case at hand, there are two 
independent modes which give two possible quantisation schemes in AdS. 
Klebanov and Witten have argued \cite{KleWit} that the existence of two 
independent modes for fields in this mass range implies the existence of 
two conformal field theories dual to the same bulk metric. These are 
called the $\D_+$ and the $\D_-$-theory. In the $\D_+$-theory, the 
lowest-order in $r$ mode, $\f_{0}$,
has the usual interpretation as an external source that couples to an 
operator $O(x)$ of conformal dimension $\D_+$, whereas $\varphi(x)$ 
(which appears at order $\D_-$) is related to the expectation value of 
$O(x)$. In the $\D_-$-theory, on the other hand, $\f_{0}$ is interpreted 
as an expectation value whereas $\varphi$ is the source. Both theories 
are related by a Legendre transformation. The case $\D_+=\D_-$ is special 
and corresponds to the tachyon of minimal mass.

Let us consider the $\D_+$-theory, where $\f_0$ corresponds to an operator 
of dimension $\D={d-1\over2}$,
\be
\bra O(x)\ket=-\f_0(x).
\ee
The expression for $\f_0$ can be obtained from \eq{45}. Notice that
as $r\rightarrow0$ the step function approaches 
$\th(u)\rightarrow\th({t^2-\vec{x}^2\over t})$. This means that the 
operator $O(x)$ has different expectation values on either side of the 
light-cone, $|t|>|\vec{x}|$ and $|t|<|\vec{x}|$, and furthermore there 
is a reflection as $t\rightarrow-t$. The operator acquires a certain 
``dressing" inside the light-cone. In the $\D_-$-theory, where $\f_0$ is 
interpreted as a source for $O(x)$, we see that the effect of the 
shock-wave is to introduce an explicit time-dependence in the source. 

As pointed out in \cite{HI} shock-waves in AdS correspond 
to states with a stress-energy tensor concentrated on the light cone. 
We have found that when we also turn on a source for an operator 
of dimension $\D={d-1\over2}$ in the background of these light-cone 
states, the operator aquires different values on either side of the 
light-cone. Elaborating this a little bit further along the lines of 
\cite{BKLT} let us add that there is a map between the creation 
and the annihilation 
operators of the field $\phi$ and the composite operators in terms of
which $O(x)$ is 
expanded. This however assumes a well-defined field theory for the 
scalar field $\phi$ in AdS, which we certainly have not constructed here 
(see however \cite{AvIsSt,BKL}). One has to find a complete set of operators 
that generate the Hilbert space of the boundary theory and that have a 
well-defined inner product. This imposes additional conditions on the 
solutions \eq{45} for them to be normalisable, like the quantisation of 
the frequencies. It would be most interesting to work out all these 
details, and to have an explicit field theory realisation of these phenomena.

The next step would be to consider gravitationally interacting fields in 
this AdS background. In reference \cite{VVexchange} it was shown that 
fields interacting by means of shock-waves on a black hole horizon satisfy 
an exchange algebra, of the form:
\be
\phi_{\tn{out}}(y)\phi_{\tn{in}}(x)=\exp\left[if^{ab}(|x-y|){\pa\over\pa x^a}
{\pa\over\pa y^b} \right] \phi_{\tn{in}}(x)\phi_{\tn{out}}(y),
\le{exchange3}
where $f^{ab}=\e^{ab}f$. Here the non-commutativity of the fields was 
ascribed to the fluctuations of the horizon due to in-coming and out-going 
shock-waves. In reference \cite{Sthesis}, an alternative derivation of this 
exchange algebra is given for Minkowski space. The derivation does not use 
the presence of a horizon, but only the fact that creation operators create 
particles that carry shock-waves with them and thus produce shifts on the 
back-ground 
space-time. This is closely related to the form \eq{45} of the solutions of
the Klein-Gordon equation, which up to a conformal factor is the same in 
AdS and in Minkowski space. Therefore it seems reasonable to expect that 
a similar kind of non-commutative behaviour is to be found in AdS.
It would be interesting to interpret this in terms of operators in the CFT. 
Notice however that when performing such a derivation one can no longer 
ignore the problem of correct quantisation of fields in AdS.

It seems likely that yet another way to derive the algebra \eq{exchange3} 
is by coupling our boundary action not to point particles but to scalar 
fields whose energy-momentum tensor is concentrated mainly in the 
longitudinal space. 

It is interesting to note that shock-wave solutions are exact solutions 
of string theory. Indeed, in \cite{AmKl2} it has been shown that shock-wave 
backgrounds are solutions to all orders in the sigma-model perturbation 
theory. In \cite{HI}, it was shown also that the AdS shock-wave does not 
receive any $\a'$-corrections using a geometrical argument
\cite{Kalloshraj,Horowitz}. The argument uses the fact that all scalar 
combinations that can be formed from the contribution to the Riemann tensor 
due to the shock-wave vanish. Thus, corrections to the supergravity action 
can only come from the AdS part of the metric, but these are known to
vanish. Thus, shock-waves are among the few known examples of exact 
backgrounds of string theory. Another interesting fact is that the 
amplitude computed by 't Hooft agrees, at large distances, with the 
amplitude of a free string in the shock-wave background generated by 
another string. So, the shock wave can 
be regarded as a non-perturbative effect coming from the resummation of 
flat-metric string contributions \cite{ACV,AmKl1}. At small distances, 
however, the string amplitudes do not exhibit the singular behaviour of 
the point particle case. Let us however point out that to our knowledge 
no amplitude valid beyond the eikonal regime has been computed so far 
for the point particle case, and so there is not much one can conclude 
from the discrepancy.

\section{Comments and conclusions}

Our analysis is a semi-classical analysis in the sense that we have 
setup a path-integral involving $S_\parallel$ that in addition involves only 
the fluctuations with insertion of fields all taking place on the boundary. 
Thus we have actually constructed a general proof of a particular form of 
holography -- that corresponding to interactions of massless particles
via gravitational shock-wave dynamics encoded in a theory of 
fluctuations on the boundary.

We would like to point out that our derivation requires no specific gauge 
choice. This agrees with \cite{VVErice} though not with the earlier paper
\cite{VV2}. However we do impose the requirement on our metric that it is 
of an approximately $2 + (d-2)$ block-diagonal form with small off-diagonal
components $h_{i\a}$. We have seen that in the course of our construction
it was indeed very important to retain the small off-diagonal 
$h_{i\a}$ as it was precisely due to this that the constraint $R_{i\a}$ was
derived and which was of importance to remove all bulk terms in the 
theory. 

The fact that in this eikonal limit the theory becomes holographic in the
sense described above is due not {\it only} to the fact that we treat
essentially as a classical background the transverse metric but also
to the crucial fact that one has an additional constraint to
impose. As already remarked this constraint arises from the
linear fluctuations in $h_{i\a}$ at order $\epsilon^{-1}$ in the rescaled
action and is therefore associated to small off-diagonal pieces of the
metric. In the end there is therefore no complete decoupling of transverse
and longitudinal components of the metric as they are 
tied together by the non-trivial constraint $R_{i\a} = 0$ \eq{constrainteq}. 

An alternative
approach to high energy scattering has been advocated by 
Amati, Ciafaloni and Veneziano \cite{ACV}. Using the regularized effective 
lagrangian proposed by Lipatov \cite{lipatov} 
they constructed an eikonal S-matrix 
that resums all semiclassical terms coming from the superstring approach. 
Their relevant gravitational degrees of freedom are the longitudinal 
modes and their ``intermediate component''. This is indeed 
reminiscent of the splitting discussed here into transverse
and longitudinal modes. The difference lies in the fact that
the intermediate component is momentum dependent (thus not purely
transverse to the incoming beam) inducing a non-locality in the
effective vertices. Complete decoupling as in the Verlinde case can be
achieved in what they call the non-interacting limit. Our analysis
suggests that the non-interacting limit cannot simply be taken 
without also taking into account precisely the off diagonal pieces.

The off-diagonal constraint essentially restricts the variations
of our solutions in the transverse directions. This is where the
dependence on the transverse direction is really taken into account. 
If follows that                            
to effectively obtain a boundary theory one simply imposes this
constraint on the transverse dynamics. We could rephrase the state of
affairs by saying that Einstein gravity in the eikonal is a topological
theory on a two-dimensional manifold embedded in $d$ dimensions, 
provided some constraints are
imposed on the ``lapse function'' $V$ which allows one to move from one plane
to another by means of transverse deformations.

A certain similarity in the imposure of constraints on the deformations
in the transverse directions  is reminiscent of Bousso's \cite{Bousso} 
idea of holographic screens. In Bousso's case the projection 
of the information is controlled by the Raychauduri
equation of classical general 
relativity. The latter describes the  focussing properties of  geodesics 
and a specific set of rules and terminology is introduced to define 
an appropriate notion of screen. In particular the Raychauduri equation
describes the 
change along light-like geodesics of the expansion $\theta$,
\be
\theta = \frac{dA}{d\lambda}
\ee
where $A$ is the cross-sectional area defined by a set of geodesics
nearby to the geodesic with affine parameter $\lambda$
\cite{HE,Wald}. Screens 
are placed at submanifolds perpendicular to the geodesics at the point
where the expansion changes from increasing to decreasing. 
These and similar considerations match indeed with the idea of 
the holographic principle realized by means of a holographic projection. 
When one examines
propagation  on  a classical background, as in our case, geodesics are not 
any more straight lines because of gravity and this has of course consequences
when one wants to project onto a boundary. The result of this 
projection in general is a non-trivial boundary theory. We would like 
to point out that 
the translation of these {\it classical} bulk fluctuations
into a projected description from the brane point of view is at 
the moment pretty much unclear  despite some recent attempts 
\cite{PolS-m,SussS-m}.

We also recall that as stressed by 't Hooft 
the gravitational interactions close
to the horizon of a black hole or more generally at high energies
are precisely described by shock wave configurations associated to
boosted particles. They have non trivial backreaction effects,
bringing about a shift in the geodesics of the outgoing particles
which induces a form of non-commutativity at the quantum level. This has 
been observed in the analysis of \cite{VVErice}  
and similarly occurs here 
in the particular subset of cases considered for which the boundary 
action is quadratic. 

Throughout the paper we have worked with the Einstein-Hilbert action 
without including higher curvature corrections. However
our method is perfectly 
applicable for these higher order terms too, and in fact considering them 
is important when the energy is increased above $1/\Pl$. 
When embedding our theory 
in a specific string theory, one also has to include additional matter 
fields. Notice, however, that for the case $d=5$ it should be straightforward 
to embed our results in string theory by condidering backgrounds with a 
constant dilaton and a covariantly constant self-dual 5-form compactified
for example on an $S^5$. This is left for future research.

\section{Acknowledgements}

We would like to thank C. Bachas, V. Balasubramanian, M.Blau, J. de Boer, 
L. Cornalba, S. Giddings, S. Itzhaki, D. Kabat, S-J. Rey, 
G. Thompson and E. and H. Verlinde for discussions, and especially 
G. 't Hooft for his continued interest and 
critical reading of the manuscript.  
G. A. warmly thanks the Spinoza Institute and 
ENS Paris for kind hospitality during this work. 
G.A. and M.O'L. are partially supported by the EEC under RTN program 
HPRN-CT-2000-00131. G.A. is also supported in part by MURST.
M.O'L. was also supported by the grant Pionier Verlinde in the 
initial stages of this work. 
S. de Haro would like to thank Soo-Jong Rey and the string theory 
group at Seoul National University for hospitality during part of this work. 
The work of SdH has been partially supported by the grant BK2000.

\appendix

\section{Scaling of curvature}

In this section we give the details of the final rescaled (up to lowest order
in $h_{i\a}$, all orders in $\e$) Ricci tensor $R_{\mu\nu}$. From here 
one can simply 
check the expansion of the Einstein-Hilbert action.
Higher orders in $h_{i\a}$ (quadratic at $1/\epsilon^2$ 
and at $\epsilon^0$) are not necessary as we are not going to consider 
the fluctuations of the metric.

\be
R_{\a\b} &\ra& \epsilon^0 ( R_{\a\b} - \frac{1}{2} \nabla_\b (g^{ik}
\partial_\a g_{ik}) - \frac{1}{4} g^{ij}\partial_\a g_{kj} 
g^{km}\partial_\b g_{im})\nn
 &+& \epsilon^2 (-\frac{1}{2} \nabla_i(g^{ij}\partial_j g_{\a\b}) 
- \frac{1}{4} g^{\g\rho}\partial_i g_{\g\rho} g^{ij}\partial_j g_{\a\b} \nn
&+& \frac{1}{4} g^{\g\rho}\partial_kg_{\b\rho} g^{ki}\partial_i g_{\a\g}
+ \frac{1}{4} g^{\g\rho}\partial_kg_{\a\rho} g^{ki}\partial_i g_{\b\g})
\ee
The leading term in $R_{i\a}$ is at zero order in $h_{i\a}$ which is already 
sufficient for our purposes as it is always multiplied by $h_{i\a}$ in the 
action and this term arises at order $\epsilon^0$
\be
R_{i\a} &\ra& \epsilon^0 (\frac{1}{2}\nabla_\b 
(g^{\b\rho}\partial_i g_{\a\rho})
 - \frac{1}{2}\nabla_\a (g^{\b\rho}\partial_i g_{\b\rho}) + \frac{1}{2}
\nabla_k(g^{kj}\partial_a g_{ij})\nn &-& \frac{1}{2} \nabla_i
(g^{kj}\partial_\a g_{kj}) + \frac{1}{4} g^{\g\rho}\partial_i g_{\a\rho} 
g^{kj} \partial_\g g_{kj} + \frac{1}{4} g^{km}\partial_\a g_{im} 
g^{\g\b} \partial_k g_{\g\b}\nn &-& \frac{1}{2} 
g^{\b\rho}\partial_\rho g_{ik} g^{kj} \partial_j g_{\a\b})
\ee
$R_{ij}$ is identical to $R_{\a\b}$ under the interchange of Greek and 
Roman indices and $\epsilon\ra\epsilon^{-1}$.
\be
R_{ij} &\ra& \epsilon^{-2} (-\frac{1}{2} \nabla_\a(g^{\a\b}\partial_\b g_{ij}) 
- \frac{1}{4} g^{km}\partial_\a g_{km} g^{\a\b}\partial_\b g_{ij} \nn
&+& \frac{1}{4} g^{km}\partial_\g g_{jm} g^{\g\a}\partial_\a g_{ik}
+ \frac{1}{4} g^{km}\partial_\g g_{im} g^{\g\a}\partial_\a g_{jk})\nn
&+& \epsilon^0 ( R_{ij} - \frac{1}{2} \nabla_j (g^{\a\g}
\partial_i g_{\a\g}) - \frac{1}{4} g^{\a\b}\partial_\a g_{\g\b} 
g^{\g\rho}\partial_j g_{\a\rho})
\ee

The exterior curvature part of the Einstein -- Hilbert action is
\be
S &=& \frac{1}{\Pl^{d-2}}\int\sqrt{K}\nabla_\mu n^\mu 
\ee
$K$ is the boundary metric which under rescaling is multiplied 
by $\ell_\parallel^2 \ell_\perp^{2(d-2)}$. The normal $n$ will have 
a non-zero component only in the direction perpendicular to the 
boundary, parallel to the longitudinal scattering plane. Thus as 
the longitudinal metric scales with $\ell_\parallel^2$ the normalisation
condition for $n$ implies that it will also scale with $\ell_\parallel$. 
Thus,
\be
\nabla_\mu n^\mu = \frac{\nabla_\a n^\a}{\ell_\parallel} + 
\frac{\nabla_i n^i}{\ell_\perp}.
\ee
The exterior curvature term of the action becomes
\be
\e^{d-4}S_{\partial M}&=& \frac{1}{\epsilon^2}\int\sqrt{K}\nabla_\a n^\a +  
\frac{1}{\epsilon}\int\sqrt{K}\nabla_i n^i.
\ee
As claimed in the text there is no additional contribution
to the boundary action coming from the exterior curvature. 

\section{Classical solutions}

In this appendix we give some more details on how to solve the equations 
of motion for the background, coming from the $\frac{1}{\epsilon^2}$ 
part of the action.

We rewrite the action \eq{Sperp} in the following form (now 
concentrating on the two-dimensional covariant part),
\be
S=-{1\over2}\int\@{-g}\left(g^{\a\b}\pa_\a\phi\pa_\b\phi+
\L\phi^2-{1\over2}\phi^2R[g]\right).
\ee
This action belongs to the class of actions considered in \cite{BOL}, 
with Lagrangian of the form
\be
L=\@{-g}\left(g^{\a\b}\pa_\a\phi\pa_\b\phi-\l\phi^{2k}-Q\phi^2R[g]\right),
\ee
with the obvious values $k=1$, $\l=-{(d-2)\over2(d-3)}\L$, 
$Q={(d-2)\over4(d-3)}$. 

As argued in the main text, we consider static metrics of the form 
\eq{longmetric}. The lagrangian (with $k=1$) then reduces to the 
particle Lagrangian
\be
L={1\over g}\left(e{\phi'}^2-4Qe'\phi\phi'\right)-\l ge\phi^2.
\ee
The prime denotes derivatives with respect to $x$. It is obvious that 
the field $g$ does not contribute to the dynamics - the equation of 
motion for $g$ is simply an expression of reparameterization invariance 
in the spatial co-ordinate. In fact, all the 
$g$-depenence disappears from the equations of motion if we define a new 
variable $r=\int_0^x\dd x'\,g(x')$. We then get
\be
-2Q\,{\ddot e\over e} +{\dot\phi\dot e\over\phi e}
+{\ddot\phi\over\phi}+\l &=&0\nn
{\ddot\phi\over\phi}+\left(1+{1\over4Q}\right)
\left({\dot\phi\over\phi}\right)^2-
{\l\over4Q}&=&0\nn
{\dot\phi\over\phi}\left({\dot\phi\over\phi}-4Q{\dot e\over e}\right)+
\l&=&0,
\le{lagreom}
and the dots denote derivatives with respect to $r$.

Substituting this solution one has for the curvature

\be
R[g]=-2a^2 \left[1+{3\gamma\over 4Q}+ {\gamma (\gamma-4Q) \over 16Q^2 } 
\left({A-Be^{-2ar}\over A+Be^{-2ar}}\right)^2    \right]
\le{curvature}
where 
\be
a^2={ \lambda \over 4Q\gamma}.
\ee
We see from \eq{curvature} that among various solutions we also have
the case in which the curvature is constant if either $A=0$ or
$B=0$.

Since both cases differ only by a coordinate transformation, we choose $B=0$. 
The longitudinal metric $g_{\alpha \beta}$ is then
\be
\dd s^2=-(aCA^{q})^2 e^{ 2aqr}\dd t^2+\dd r^2.
\ee
where
\be
q=1+{ \gamma \over 4Q}
\ee

This is indeed the $AdS_2$ metric with the proper warp factor growing 
linearly in the radial coordinate. Our coordinates, however, do not 
cover the whole of AdS. One finds global coordinates by defining
\be
e^{aqr}=\cos\rho,
\ee
where $0\leq\rho\leq\pi/2$.

The curvature $R[g]=-{2\ddot e\over e}$ obviously simplifies and becomes
\be
R[g]=-{\lambda(4Q+\gamma) \over 8 Q^2 \gamma }
\ee
Furthermore, those solutions with $A$ and $B$ non-zero will be analogous 
in structure 
to $AdS_2$/Schwarzschild geometries, though the metric will have
a different functional form due to the presence of the non-trivial scalar 
field. The mass of the configuration will be proportional to $B$. 

In $d=3$ there are small modifications due to the appearence of several
$(d-3)$ factors in the general solutions. We can easily proceed here as 
follows. The one-dimensional form of the action is:
\be
L={e'{\phi^2}'\over g}-\L eg\phi^2,
\ee
and so the equations of motion reduce to
\be
\ddot\phi^2+\L\phi^2&=&0\nn
\ddot e+\L e&=&0\nn
{\dot\phi^2\dot e\over\phi^2 e}+\L&=&0,
\ee 
after reabsorbing the non-dynamical field $g$ in the definition 
of the parameter $r$, as before. 

\subsection{Global structure of the solutions}

The metric
\be
\dd s^2=-e(r)^2\dd t^2+\dd r^2
\ee
has a horizon when $e(r)=0$. There are two possible locations of this 
horizon, depending on the relative sign of the initial conditions $A$ and $B$.

For $B/A>0$, $e(r)$ has a simple zero. With the following rescalings 
of the coordinates,
\be
r&=&\sqrt{Q\g\over\l}\log B/A+\et\nn
t&=&{4Q\g\over C\l}(4AB)^{-\g/8Q-1/2}\tau
\ee
the metric near the horizon  is simply the Rindler space metric,
\be
\dd s^2=-\et^2\dd\tau^2+\dd\et^2,
\le{Rindler}
and so locally the space is flat.

For $B/A<0$, we rescale the coordinates as follows:
\be
r&=&\sqrt{Q\g\over\l}\log|B/A|+\et\nn
t&=&({\l|AB|\over Q\g})^{-\g/8Q-1/2}\tau,
\ee
and we find the metric
\be
\dd s^2=-\et^{\g/2Q}\dd\tau^2+\dd\et^2
\le{metric2}
with curvature $R=-{\g(\g-4Q)\over8Q^2\et^2}$.

\section{Null geodesics in AdS and the stress-energy tensor}
\label{nullgeodesics}

The stress-energy tensor induced by a massive particle travelling along a 
path $\g$ described by the trajectory $z^\m(s)$ is given by
\be
T^{\m\n}(x)=-\frac{p}{\@{-G(x)}}\int_\g\dd 
s\,\d^d(x-z(s))\,\dot{z}^\m\dot{z}^\n,
\le{6}
with the usual normalisation for delta-functions. $p$ is the momentum of 
the particle along the light-cone.

In order to be able to use \eq{6} for practical purposes, one has to 
compute the trajectories $z^\m(s)$ with some boundary conditions. We will 
concentrate on trajectories of particles the momentum of which has components 
only in the $v$-direction, but it is a straightforward exercise to consider 
other cases.

It is well-known that the null geodesics of two conformally related 
space-times are the same up to a reparametrisation of the geodesic length. 
Therefore null trajectories in the above co-ordinates will take the same 
form as those in Minkowski space. It is nevertheless convenient for the 
computation of the stress-energy tensor to see explicitly how the affine 
parameter changes.

The geodesic equation and the mass-shell condition give
\be
{\dd\over\dd\l}\left({\et_{\m\n}\dot z^\n\over\O^2}\right) &=& 
2{\et_{\m\n}z^\n{\cal L}\over\ell^2\,\O}\nn
{\cal L}&=&{1\over\O^2}\,\et_{\m\n}\dot z^\m\dot z^\n=0.
\le{A11b}
${\cal L}$ is the Lagrange density defined by the second of \eq{A11b},
and $\l$ the affine parameter along the geodesic. These equations integrate to
\be
\et_{\m\n}\,\dot z^\n&=&v_\m\,\O^2.
\le{A12}
$v_\m$ is a constant, lightlike vector satisfying $\et^{\m\n}v_\m v_\n=0$ 
to be determined by the boundary conditions. This equation also relates 
the affine parameter in AdS to the affine parameter in Minkowski space.

The stress-energy tensor of this particle now equals:
\be
T_{\m\n}&=&-p\,\O^d\,v_\m v_\n\int\dd s\,\d^d(y-z(s)),
\le{A13}
and choosing co-ordinates where momentum is purely in the $v$-direction, 
this reduces to
\be
T_{uu}&=&-p\,\O^d\,\d(u-u_0)\,\d(\r-\r_0),
\le{A15}
where $\r=\sum_{i=1}^{d-2}y_i^2$. Notice that in order for the metric 
\eq{HI} to be a solution of Einstein's equations with this stress-energy 
tensor we need the initial condition $u_0=0$. It is also convenient to 
take $\r_0=0$. Thus we get the stress-energy tensor,
\be
T_{uu}=-p\d(u)\d(\r).
\le{15}

Once one has \eq{15}, one can compute the back-reaction on the AdS metric, 
obtaining the solution found by Horowitz and Itzhaki. The next step is then 
to compute the geodesics of a test particle in the back-reaction corrected 
metric. The computation goes along the same lines as the one above.
We do not give the details here since it is a 
straightforward exercise, but give only the results. We concentrate on 
trajectories whose initial velocities are perpendicular to the velocity of 
the shockwave, that is, the geodesics with $v=y^i=0$ before the collision. 
This gives a head-on collision.

It turns out that the geodesic equations can again be exactly integrated, 
and the effect is the same as in Minkowski space: there is a shift in the 
$v$ coordinate and a deflection in the $x^i$-plane which nevertheless is 
negligible in the eikonal approximation where the impact parameter is much 
larger than the Planck length. In this approximation, the shift is given by
\be
\d v=-8\pi\GN\,p_u\,F_0\,\th(u),
\ee
where $F_0$ is the shift function before the collision, $F_0=F(u=0)$.

Of course the same results can be found from geodesics in Minkowski space by 
noting that massless geodesics are invariant under conformal transformations 
of the metric.

It is interesting to note that, when one considers only one particle, there 
is no self-interaction, and therefore the present solution to the 
Einstein-matter system with the given boundary conditions is exact. However, 
when considering two particles this is not true anymore, and one has to 
restrict oneself to consider a ``soft" test particle in the background 
of a ``hard" particle.

\end{document}